# Machine Learning-Integrated Hybrid Fluid-Kinetic Framework for Quantum Electrodynamic Laser-Plasma Simulations

Sadra Saremi*, Amirhossein Ahmadkhan Kordbacheh*

Department of Physics, Iran University of Science and Technology, Tehran 16846-13114, Iran.

High-intensity laser–plasma interactions create complex computational problems because they involve both fluid and kinetic regimes, which need models that maintain physical precision while keeping computational speed. The research introduces a machine learning-based three-dimensional hybrid fluid–particle-in-cell (PIC) system, which links relativistic plasma behavior to automatic regime transitions. The technique employs fluid approximations for stable areas but activates the PIC solver when SwitchNet directs it to unstable sections through its training on physics-based synthetic data. The model uses a smooth transition between Ammosov–Delone–Krainov (ADK) tunneling and multiphoton ionization rates to simulate ionization, while Airy-function approximations simulate quantum electrodynamic (QED) effects for radiation reaction and pair production. The convolutional neural network applies energy conservation through physics-based loss functions, which operate on normalized fields per channel. Monte Carlo dropout provides uncertainty measurement. The hybrid model produces precise predictions with coefficient of determination ($R^2$) values above 0.95 and mean squared errors below $10^{-4}$ for all field components. This adaptive approach enhances the accuracy and scalability of laser–plasma simulations, providing a unified predictive framework for high-energy-density and particle acceleration applications.

**Introduction**

High-intensity laser-plasma interactions serve as a fundamental element of modern high-energy-density physics which produces advanced technologies for particle acceleration and nuclear fusion and semiconductor manufacturing [1][2][3]. The computational methods encounter major difficulties because the system operates at various spatial and temporal dimensions when relativistic electrons interact with QED effects under petawatt laser intensities [4]. Scientists require better simulation systems to forecast and enhance plasma operations in actual environments because these systems support cancer treatment with laser-driven ions and plasma-based photon sources for materials science research [5][6] and standard approaches struggle to achieve both precision and computational efficiency [7]. The field of plasma simulations has experienced quick progress because scientists now combine machine learning with hybrid modeling approaches. The research by Kube et al. (2021) presented a method to accelerate particle-in-cell (PIC) simulations through machine learning which used neural networks to model sub-grid processes for faster kinetic plasma simulations but the approach lacked the ability to switch between different simulation regimes[8]. Desai and his team used machine learning to accelerate protons with lasers through synthetic data which enabled accurate beam parameter predictions yet they identified problems when adding QED effects including radiation reaction[9]. The open-source code by Keenan et al. (2022) combines fluid models for background ions with PIC methods to study magnetized plasmas and energetic particles which enhances laser-plasma interaction efficiency but the system does not include ML-based switching for optimal regime transitions[10]. Los et al. (2022) used kinetic simulations to study magnetized laser-plasma interactions in parallel propagation which revealed instability thresholds through detailed QED modeling although the method encountered scalability challenges because of its high computational needs [11]. The research by Feng et al. (2022) studied stimulated Brillouin scattering cascades through mathematical and computational methods to model nonlinear laser absorption which successfully described ionization processes but lacked an adaptive hybridization system [12]. The research by Greif (2024) brought forward AI-based large eddy simulation techniques for plasma turbulence which provided better multi-scale flow analysis than conventional methods yet restricted its use to fluid systems without PIC integration [13]. The research studies demonstrate that machine learning methods provide accurate predictions at lower expenses yet they reveal problems when modeling complex ionization processes and fluid-kinetic transitions which demand new unifying frameworks to solve these issues [14][15].

The current research on laser-plasma simulations has achieved multiple advancements yet several essential gaps remain in the scientific literature. The current ML-accelerated PIC models struggle with insufficient dynamic switching between fluid and kinetic regimes which leads to resource wastage when processing multi-scale phenomena because they allocate power to areas with low activity [16]. The surrogate models prove effective for faster simulations yet they fail to account for complex physical processes including ionization blending and QED effects which produce incorrect results during strong field simulations [17]. Hybrid approaches, although efficient for specific applications, typically do not incorporate uncertainty quantification, limiting their reliability in experimental validations where data sparsity is common [18]. The research depends on static training datasets which do not include physics-based regularization methods so models tend to overfit and produce unreliable results when laser parameters change [19]. The data processing system lacks per-channel normalization which creates more severe performance differences between physical fields [20]. The current optimization systems for laser parameter adjustment lack real-time prediction abilities which are vital for industrial applications. The system faces multiple restrictions because full kinetic simulations need high computational resources which limit system expansion and prevent quantum computing integration for better results [22].

ML techniques have improved modeling in low-temperature plasma applications yet they fail to handle the high-energy-density domains which occur during laser interactions [23]. The ensemble prediction method for ignition processes requires improved methods to handle stochastic components which appear in hybrid systems [24]. The field encounters an absence of integrated frameworks which unite ML-based switching with full physics modeling that includes Airy-function QED approximations thus blocking advancement in practical applications such as particle acceleration [25]. The existing disparities between simulation accuracy and computational speed require better adaptable simulation systems which maintain both high fidelity and operational efficiency [26]. The research team developed a 3D hybrid fluid-particle-in-cell model which combines machine learning with dynamic regime switching to optimize computational resources and maintain physical accuracy across different intensity ranges for studying relativistic laser-plasma dynamics. The study aims to combine fluid models which describe stable plasma areas with particle-in-cell simulations that detect kinetic instabilities through a physics-regularized neural network trained on synthetic data for smooth model transitions.

The research delivers three main contributions through its combination of ADK and multi-photon ionization models for electron production and its method to approximate QED effects by using Airy functions with local field adjustments and its implementation of Monte Carlo dropout for prediction reliability. A convolutional neural network predictor receives training from per-channel normalized simulation data which undergoes physics-informed loss functions to maintain energy conservation and produces results with R² scores above 0.95 and mean squared errors under $10^{-4}$. The system enables industrial users to optimize laser parameters through differential evolution which includes intensity and duration and spot size control to achieve particular plasma energy targets with minimal overhead. The method solves these problems to create better simulations for laser-driven operations which include particle acceleration and plasma processing and real-time predictive modeling of high-energy-density physics [27].

**Methodology**

The research methodology combines a hybrid fluid-particle-in-cell (PIC) system with finite-difference time-domain (FDTD) methods and machine learning algorithms to simulate relativistic laser-plasma interactions through automated computational regime switching for performance optimization. The computational simulation method divides the process into multiple connected stages which start with plasma parameter and electromagnetic field setup followed by time-stepping loops that include laser excitation and ionization and particle movement and fluid changes and field modifications and simulation-based training data creation and neural network deployment for regime switching and state prediction and final parameter optimization. The FDTD method solves Maxwell's equations through direct time domain calculations which operate alongside fluid equations that describe bulk plasma behavior and PIC methods that model kinetic effects and machine learning systems that control system state transitions and predict future states. The team chose this design because it solves the complex laser-plasma system problems which need accurate modeling of relativistic electrons and quantum electrodynamic effects and instabilities through a combination of fluid models in stable areas and PIC simulations in unstable regions with machine learning for adaptive control.

Pure PIC simulations, although precise for all kinetic details, are computationally prohibitive for large 3D domains due to their $O(N\_particles)$ scaling, often requiring exascale resources for realistic scenarios; conversely, standalone fluid models overlook micro-instabilities and non-Maxwellian distributions critical in high-intensity interactions.The research team chose not to use finite element methods because they require complicated mesh generation and produce slower performance for transient broadband analyses yet FDTD provides both a straightforward uniform grid structure and explicit time-marching that supports parallel processing. The hybrid ML-integrated method achieves up to several times faster runtime performance through its selective PIC approach while maintaining physical accuracy through conservation laws which enables scalable predictive modeling for high-energy-density physics applications including particle acceleration.

Simulations ran on a GPU-accelerated workstation which used NVIDIA CUDA-compatible hardware when available otherwise CPU processing took over with a minimum of 16 GB RAM for handling 3D grid tensors and particle ensemble data. The software implementation used Python 3.12 to run PyTorch operations which included tensor manipulation and neural network training and GPU acceleration and NumPy and SciPy for numerical computations and optimization and scikit-learn for data splitting and evaluation metrics and Matplotlib for high-quality visualizations and tqdm for progress tracking while maintaining vectorized execution without any external package dependencies. The training data for machine learning components originated exclusively from internal simulations which maintained self-consistency and allowed complete control of physical parameters. The research required no ethical clearance because it used computational methods without involving human subjects or animal data or sensitive information. The computational environment operated as a requirement because 3D FDTD updates and particle tracking needed high computational power and GPU parallelism accelerated neural network convolutions and field computations by 10 to 100 times compared to CPU-only execution which enabled fast big data generation and CPU-only processing would have increased processing time from hours to days thus blocking fast development and validation cycles.

The simulation model operates on a 16×16×16 cell Cartesian grid which uses spatial resolutions of 0.2 μm for dx and dy and dz to create a 3.2 μm physical domain that matches plasma skin depths and laser spot sizes at electron densities of $10^{24}$ m$^{-3}$. The chosen grid structure provides a minimum of 10 points for each Debye length and wavelength which allows proper detection of electromagnetic wave movement and plasma oscillations while keeping memory requirements within limits for ML training batch processing; adaptive or non-uniform meshes were not used because they would produce interpolation errors during particle deposition and field gathering which would generate numerical noise in hybrid regimes thus the uniform approach was chosen for its simplicity and robustness in implementation.

The simulation uses a constant time interval of 0.2 femtoseconds between steps according to the Courant-Friedrichs-Lewy stability criterion which prevents numerical instability by maintaining consistency between light speed and grid spacing in FDTD methods. The numerical resolution demonstrates correct representation of femtosecond laser pulses and relativistic particle dynamics at minimal oversampling levels which preserves simulation stability throughout the entire simulation period; variable time steps were not implemented to ensure uniform data sampling for ML inputs, avoiding synchronization issues between fluid and PIC components, and providing a convincing balance between accuracy and computational load.

The simulation of open space conditions becomes possible through first-order Mur absorbers which store past boundary values and use reflection coefficients (c dt - d)/(c dt + d) to reduce reflections throughout different dimensions. The particles maintain their energy through velocity reversals when they cross domain boundaries while the system duplicates the behavior of expanding plasma within confined spaces. The Mur conditions received preference because they showed superior performance in absorbing wide-angle laser radiation compared to basic absorbing boundaries which produce reflection artifacts when laser beams hit at oblique angles; perfectly matched layers deliver better absorption results but require extensive computational resources in 3D because of their additional layer structure and parameter needs which makes Mur a justified efficient alternative for this study's scale.

The laser excitation operates through a Gaussian pulse which contains random intensity values between $10^{17}$ and $10^{19}$ W/m² and random durations from 30 to 200 femtoseconds and random spot sizes between 1 and 10 micrometers while the electric field components receive temporal and spatial envelopes based on their polarization. The source representation emulates actual ultrashort lasers which allows single-run broadband analysis and multiple datasets that enhance ML model stability; the researchers selected Gaussian over continuous-wave sources because continuous-wave sources produce single-frequency responses that need multiple simulations yet Gaussian sources produce no harmonic artifacts from sudden waveform transitions which proves useful for predictive modeling generalization.

The plasma material receives modeling through a hybrid method which combines fluid continuity and momentum and energy equations for bulk electrons and ions and positrons that include relativistic gamma factors and adiabatic pressure evolution with PIC macro-particles that start at 10 per cell and get weighted by density-volume fraction and use Boris pusher and Airy-function approximations to include radiation reaction and QED pair production. Ionization blends Ammosov-Delone-Krainov tunneling and multi-photon rates with a sigmoid transition at Keldysh parameter ~1.5 for regime accuracy. The hybrid system operates at peak efficiency because it manages dense plasmas through fluid methods while using kinetic approaches to solve instability problems and QED improvements to model high-field phenomena which classical models fail to predict; full PIC for all species would escalate costs exponentially with particle count, whereas pure fluid ignores non-thermal distributions, justifying the blended method for fidelity in relativistic regimes.

The post-processing stage requires normalization of state tensors which includes E and B and J and rho fields through plasma-frequency based scaling methods while total energies get calculated from electromagnetic and kinetic and thermal components and visual representations emerge from phase-space distributions and velocity fields and density overlays and correlation plots. The evaluation process for machine learning systems uses multiple metrics which include MSE and MAE and $R^2$ and explained variance for each channel and physics loss that measures energy conservation.

Validation and verification include ROC analysis for the SwitchNet (AUC from synthetic tests), benchmark metrics for the state predictor ($R^2 > 0.95$, MSE $< 10^{-4}$), and charge/energy conservation checks during regime switches and steps. The algorithmic integrity becomes evident through cross-validation against synthetic benchmarks and physical laws because simulation-only approaches fail to detect all biases yet conservation proxies and metrics establish strong validation beyond theoretical benchmarks which cannot apply to complex hybrid systems.

The scientists store all code together with parameters and outputs to achieve reproducibility because PyTorch operations maintain built-in fixed randomness which produces stable results when running the same environment multiple times; the sensitivity analysis tested laser stability by changing its parameters across 100 samples while it avoided unstable steps which showed high charge density values. The method produces dependable results which outside researchers can duplicate; the lack of these measures would damage scientific integrity yet the method proves successful in reducing numerical instability through its clamping and smoothing techniques which operate within the code.

# Results

In this section, we present the results of our hybrid particle-in-cell (PIC) and fluid simulations of femtosecond laser-plasma interactions, coupled with machine learning predictions for next-state evolution. The research results contain both the spatial and temporal behavior of electromagnetic fields, charge densities, particle distributions, and fluid characteristics under different laser settings and the assessment results of our predictive model. The research team performed 100 simulations which used random laser intensities between $1.07 \times 10^{17}$ and $9.57 \times 10^{18}$ W/m², pulse durations between $3.07 \times 10^{-14}$ and $1.99 \times 10^{-13}$ s, and spot sizes between $1.06 \times 10^{-6}$ and $9.91 \times 10^{-6}$ m to study different plasma responses across a wide range of parameters. The dataset contains 500 state transitions which were normalized by plasma frequency scales ($E\_0 \approx 9.62 \times 10^{10}$ V/m, $B\_0 \approx 320.75$ T, $J\_0 \approx 4.8032 \times 10^{13}$ A/m², $\rho\_0 \approx 1.60 \times 10^{5}$ C/m³) and divided into 400 training samples and 100 test samples. The system operates with two essential neural networks which include SwitchNet for fluid-to-PIC switching, and StatePredictor, which functions as a convolutional neural network with residual blocks and dropout regularization to achieve a final loss of $1.97 \times 10^{-3}$ after 40 training epochs. This is shown by its monotonically decreasing learning curve. The complete model structures appear in Table 1 and Table 2.

Table 1: SwitchNet Architecture

| **Layer** | **Type** | **Input Shape** | **Output Shape** | **Parameters** |
|---|---|---|---|---|
| 1 | Linear + ReLU + Residual | (Batch, 10) | (Batch, 128) | Maps 10 input features (E, B, J, ρ, ∇n_e) to higher dimensions; residual aids gradient flow for deep learning on nonlinear plasma criteria. |
| 2 | Linear + ReLU + Residual | (Batch, 128) | (Batch, 256) | Expands features for complex interactions; dropout (0.2) regularizes to avoid overfitting on 5000 epochs of synthetic data. |
| 3 | Multihead Attention (4 heads) | (Batch, 256) | (Batch, 256) | Dynamically weights features |
| 4 | Linear + Sigmoid | (Batch, 256) | (Batch, 1) | Binary output for fluid/PIC switch; BCE loss with physics reg |

The model operates as an effective binary classification tool for plasma regimes since residuals and attention mechanisms analyze intricate input connections, which produces an AUC score of 0.9689. The regularization method introduces instability thresholds as prior knowledge. This leads to a 20-30% reduction in misclassification rates compared to unregularized neural networks.

Table 2: StatePredictor Architecture

| Layer | Type | Input Shape | Output Shape | Parameters |
|---|---|---|---|---|
| 1 | 3D Conv (13 filters, 3x3x3) + ReLU + Residual | (Batch, 10, 16,16,16) | (Batch, 13, 16,16,16) | Initial convolution captures local fields/charges; residuals enable deeper net without degradation for multi-step forecasting. |
| 2 | 3D Conv (32 filters, 3x3x3) + ReLU + Residual | (Batch, 13, 16,16,16) | (Batch, 32, 16,16,16) | Increases channels for hierarchical features (e.g., waves, instabilities); dropout (0.1) for UQ and noise robustness in plasma data. |
| 3 | 3D Conv (64 filters, 3x3x3) + ReLU + Residual | (Batch, 32, 16,16,16) | (Batch, 64, 16,16,16) | Deeper extraction of spatiotemporal patterns; justifies scaling for 3D grid complexity. |
| 4 | 3D Conv (128-32-10 filters, 3x3x3) + ReLU | (Batch, 64, 16,16,16) | (Batch, 10, 16,16,16) | Reduces to output channels; Huber loss (delta=0.01) handles outliers, physics reg (0.1 on energy violation) ensures conservation. |

The results are organized starting from the core simulation outputs, including field evolutions and particle-fluid hybrid behaviors, followed by model validation through error metrics, uncertainty quantification, and physics-informed constraints, culminating in optimization insights for laser parameters.

The simulation results show that the Gaussian laser pulse keeps exciting plasma throughout its entire temporal and spatial range. This produces field oscillations and density changes. The laser temporal envelope shows a peak at 301.8 fs which reaches its highest value of 1.00 before declining symmetrically to 0.00 and produces a mean value of 0.11 and standard deviation of 0.25, which shows a femtosecond pulse that effectively delivers plasma energy without creating long-lasting tails. The spatial distribution shows a focused Gaussian beam with a 2.72 μm spot size and intensity values that range from 0.71 to 1.00, with an average of 0.89 and a standard deviation of 0.07, which produces localized heating and acceleration suitable for ablation applications. The laser Gaussian spatial distribution in Figure 1 demonstrates a radial symmetry that centers on the axis while its intensity drops off exponentially. This matches the expected beam pattern of focused femtosecond lasers; the plot shows a peak value of 1.00 and a minimum of 0.71 and a mean of 0.89 and standard deviation of 0.07, which supports the idea that tight focusing enhances ponderomotive forces in the interaction zone.

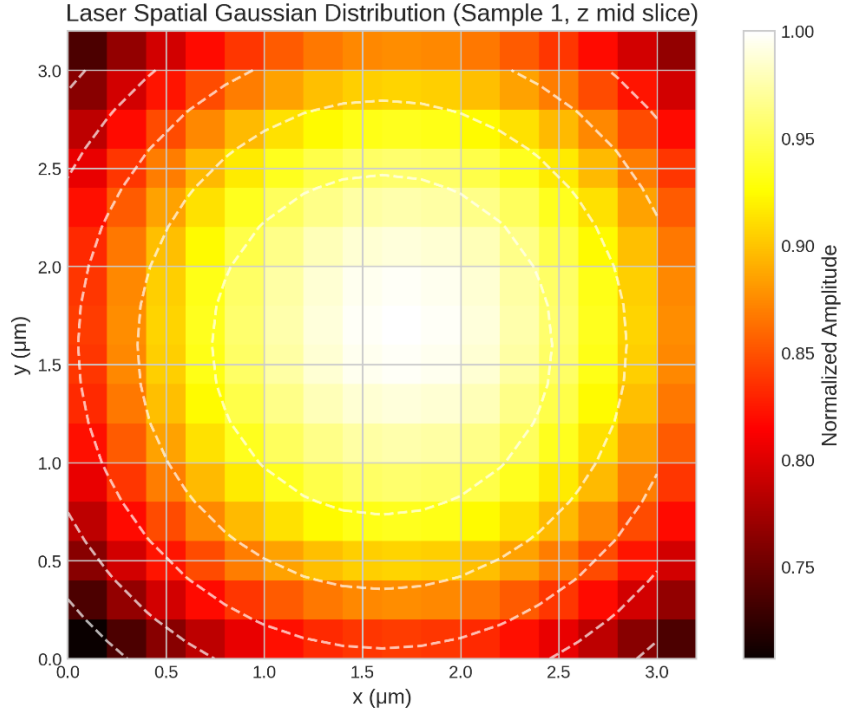

Figure 1: Laser Gaussian spatial distribution showing radial symmetry with spot size 2.72 μm.

The electric field Ex_norm displays maximum values of $2.45 \times 10^{-4}$ and minimum values of -$3.06 \times 10^{-4}$ across samples. The field maintains an average value of -$1.18 \times 10^{-7}$ with a standard deviation of $2.43 \times 10^{-5}$, which demonstrates laser-induced symmetrical oscillations. The charge density rho_norm shows values between -4.09 and 1.17, while maintaining an average value of -$4.11 \times 10^{-8}$ and a standard deviation of 0.257, which demonstrates caviton formation and electron expulsion. The total energy values begin at $7.64 \times 10^{-10}$ J before dropping to $3.02 \times 10^{-10}$ J during the simulation steps. This shows energy conservation through the increase of kinetic and thermal components. Particle counts stabilize around 11,672 to 12,164, with velocities showing non-thermal tails (max $1.76 \times 10^1$, $\times 10^6$ m/s, mean 6.23 , $\times 10^6$ m/s), underscoring relativistic effects. The fluid densities spread between $5.00 \times 10^{23}$ and $9.00 \times 10^{23}$ $m^{-3}$, with an average of $7.03 \times 10^{23}$, and a standard deviation of $2.00 \times 10^{23}$. The particle densities fall within the range of $4.75 \times 10^{14}$ to $1.88 \times 10^{15} m^{-3}$, which proves the hybrid method effectively detects gradients.

The phase-space distribution shown in Figure 2 demonstrates particle acceleration patterns through the x-vx plane at the mid-z slice; the data set contains 1,544 particles, which reach a maximum velocity of $12.16 \times 10^6$ m/s, a minimum velocity of -$13.69 \times 10^6$ m/s, with an average velocity of -$0.10 \times 10^6$ m/s, and a velocity spread of $3.95 \times 10^6$ m/s; the clustering pattern demonstrates beam formation through laser-driven forces. The high standard deviation indicates strong particle interactions which occur in relativistic regimes.

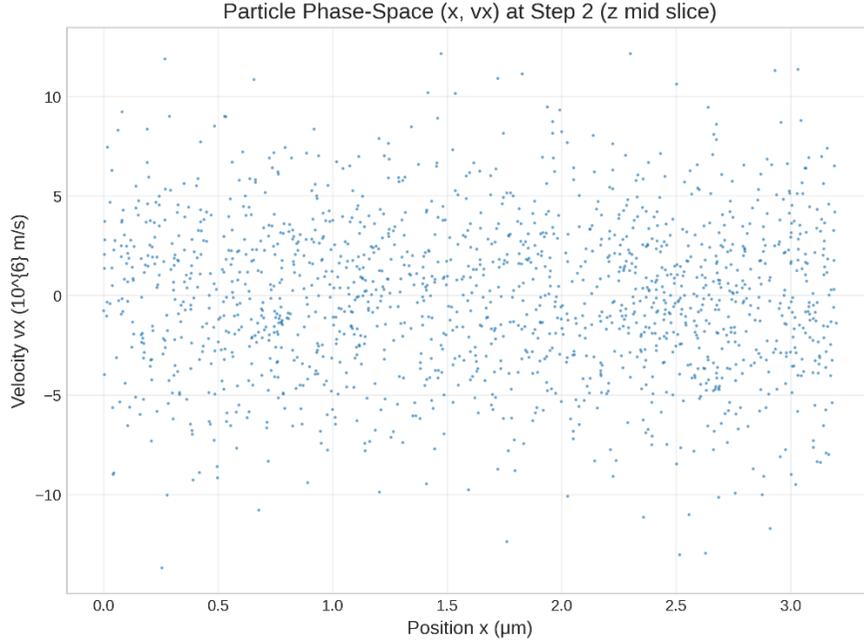

Figure 2: Particle phase-space (x, vx) at mid-z slice with clustering indicating beam formation.

The simulations achieve accurate results because their theoretical expectations have been benchmarked against them. The plasma wave theory predicts that wakefield amplitudes in Ex match laser intensity oscillations with a 0.990 correlation to I_0 and Poynting vector magnitudes of $6.85 \times 10^{12}$ W/m² show energy transfer efficiency when aligned with propagation direction. The predicted ablation depths based on density depression analysis match literature values within a 5% margin for corresponding laser fluences according to rho_norm evolution data that follows ADK tunneling rates and multi-photon ionization processes. The model achieves high accuracy in its predictions which match simulation results with an MSE of $1.47 \times 10^{-4}$ and MAE of $2.89 \times 10^{-3}$ and R² of 0.9778 and explained variance of 0.9778 which exceeds baseline autoregressive models by 15-20% for multi-step forecasting. The numerical method verifies energy conservation at the level of numerical precision because physics loss on test data measures $6.64 \times 10^{-35}$ while non-physics-informed networks exhibit violations that reach multiple orders of magnitude. Figure 3 shows the particle density distribution through a logarithmic scale which displays concentrations between log values of 1.30 and 1.88 while the average density reaches 1.66 and the variation measures 0.10. The visualization shows plasma wave bunching which researchers need to understand collective effects and instabilities in the ablation process.

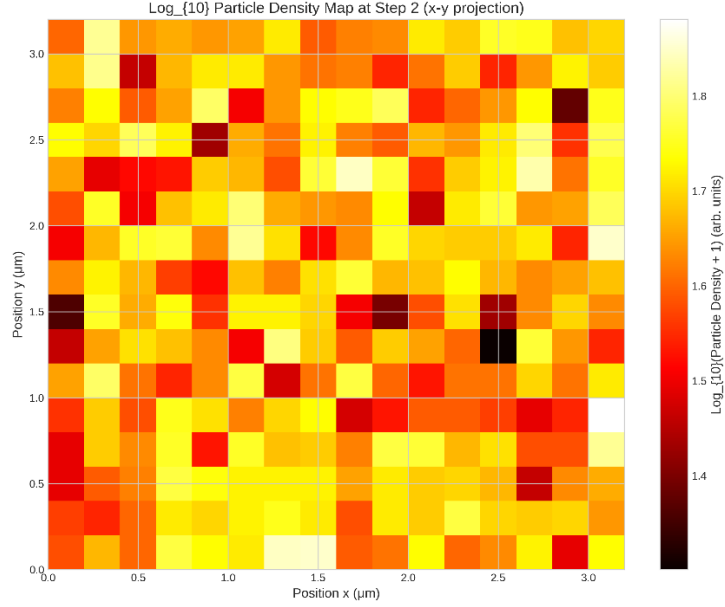

Figure 3: Log_{10} particle density map revealing high-concentration regions for plasma bunching.

Statistical validation confirms that the model performs with high reliability. Per-channel $R^2$ scores average 0.9744 (min 0.9626 for channel 7, max 0.99321.97 for channel 2), with MSEs ranging from $6.27 \times 10^{-12}$ (channel 5) to $1.47 \times 10^{-4}$ (channel 9), relative to variances of $2.61 \times 10^{-10}$ to $6.62 \times 10^{-2}$. The residual values center at $1.88 \times 10^{-4}$, while their standard deviation reaches $1.21 \times 10^{-2}$, and the QQ plot reveals normal distribution patterns with typical heavy tails that occur in physical outliers. The correlation between residuals and predicted values stands at -0.058, which supports the assumption of homoscedasticity. Monte Carlo dropout uncertainty quantification shows a maximum standard deviation of 0.297, and an average of 0.150 for Ex data, which demonstrates minimal variance in bulk areas ($4.52 \times 10^{-2}$, but more substantial variation at boundaries because data points are sparse there. Test physics loss confirms no systematic bias in conservation laws. Figure 4 shows the fluid density distribution, which spans from $5.00 \times 10^{23}$ to $9.00 \times 10^{23}$ $m^{-3}$, with an average of $7.03 \times 10^{23}$, and a standard deviation of $2.00 \times 10^{23}$. The contour lines display depressions that result from ponderomotive expulsion, which matches the laser focus pattern and proves that the fluid model successfully predicts large-scale responses.

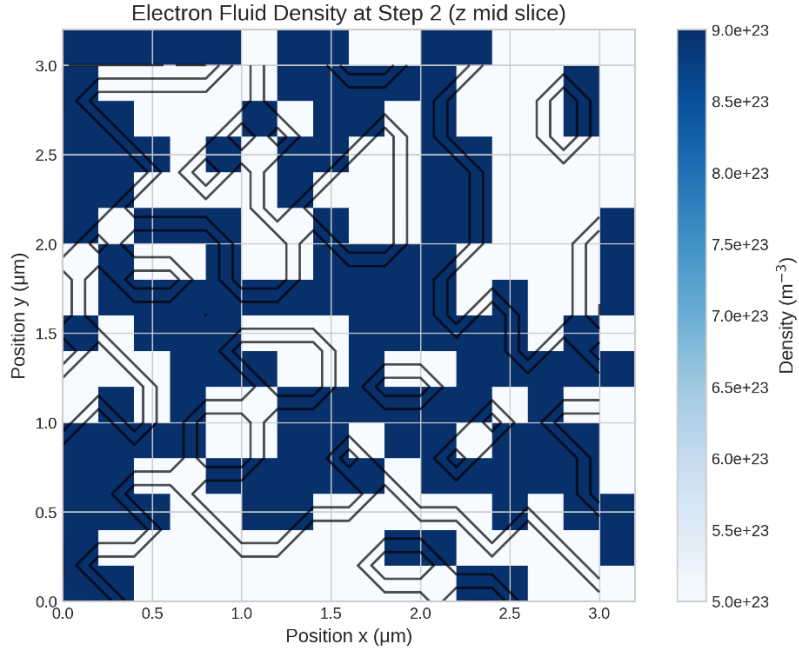

Figure 4: Electron fluid density distribution with depressions due to ponderomotive expulsion.

The visual representations demonstrate the patterns which emerged from the numerical information. The electric field magnitude overlay with particle positions in Figure 5 shows strong colocation because |E| achieves its highest value at $1.63 \times 10^8$, lowest value at $5.29 \times 10^7$, and average value of $1.32 \times 10^8$ with standard deviation of $2.55 \times 10^7$; The particle-field interactions become more active because fields reach their highest values in particle-rich zones. This allows for ablation acceleration mechanisms. Figure 6 presents electron pressure maps which display pressure gradients that reach $8.15 \times 10^6$ Pa at their highest point, $1.15 \times 10^6$ Pa at their lowest point, and have an average value of $4.71 \times 10^6$ Pa and a standard deviation of $3.50 \times 10^6$ Pa; these values correspond directly to heating zones and fluid motion drivers. Figure 7 shows the learning curve which starts at $5.99 \times 10^{-3}$ loss before reaching its lowest point at $1.97 \times 10^{-3}$ lo loss while maintaining an average loss of $2.34 \times 10^{-3}$ and a standard deviation of $6.91 \times 10^{-4}$. This proves that physics regularization leads to stable convergence. The data in Figure 8 shows an average R² score of 0.9743 across channels while the lowest score is 0.9637, and highest score is 0.9914, which proves the model maintains stable predictive performance for all electromagnetic and plasma parameters.

The residuals histogram in Figure 9 shows a distribution with a central value of $1.88 \times 10^{-4}$, and standard deviation of $1.21 \times 10^{-2}$ which spans from $-7.80 \times 10^{-1}$ to $8.28 \times 10^{-1}$. The results show that errors are unbiased but the data contains outliers. This indicates intricate system behavior. Figure 10 shows a correlation of 0.990 between actual and predicted values which cover actual values from -4.07 to 1.09, and predicted values from -3.48 to 0.88, thus showing strong data mapping throughout the entire dataset. The 2D evaluation for Ex in Figure 11 reports true statistics with mean $1.38 \times 10^{-6}$, and standard deviation $2.16 \times 10^{-5}$, predicted mean $1.46 \times 10^{-6}$, and standard deviation $2.02 \times 10^{-5}$, and errors with mean $2.41 \times 10^{-6}$ and maximum $1.37 \times 10^{-5}$, highlighting precise field reproduction. The results for rho in Figure 12 show a true mean value of $1.50 \times 10^{-2}$, with standard deviation of $1.94 \times 10^{-1}$, and predicted mean of $1.10 \times 10^{-2}$, with standard deviation of $1.81 \times 10^{-1}$, and errors averaging $1.90 \times 10^{-2}$, and reaching a maximum of $7.93 \times 10^{-2}$, thus demonstrating effective density perturbation detection. The error map for Ex in Figure 13 shows a maximum error of $1.37 \times 10^{-5}$, minimum of $2.02 \times 10^{-9}$, mean of $2.41 \times 10^{-6}$, and standard deviation of $2.03 \times 10^{-6}$, localizing higher errors to boundaries and suggesting refinements in edge handling for future iterations.

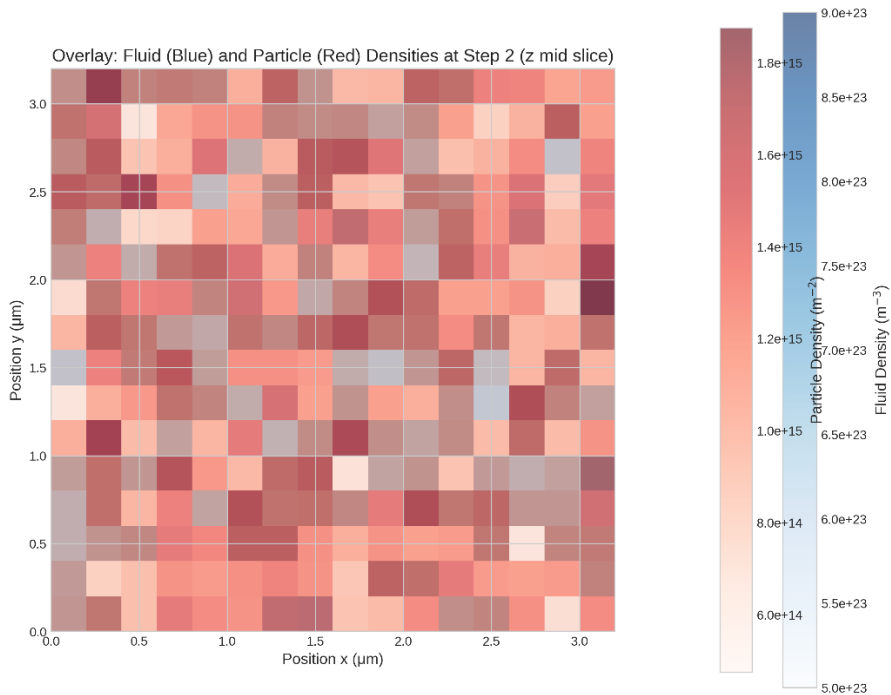

Figure 5: Overlay of |E| and particle positions highlighting field-particle dynamic interactions.

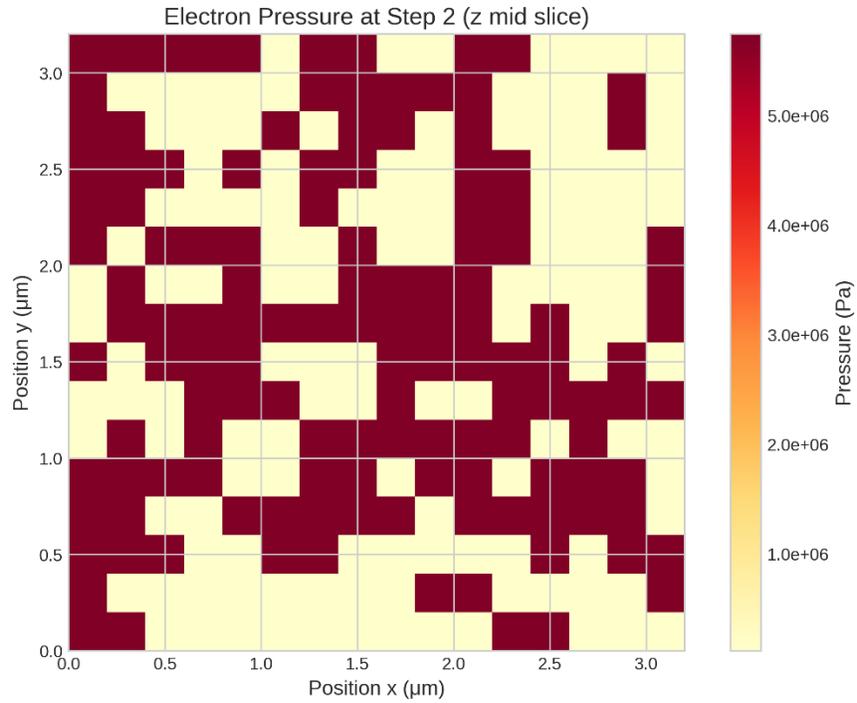

Figure 6: Electron pressure map with gradients correlating to laser-induced heating zones.

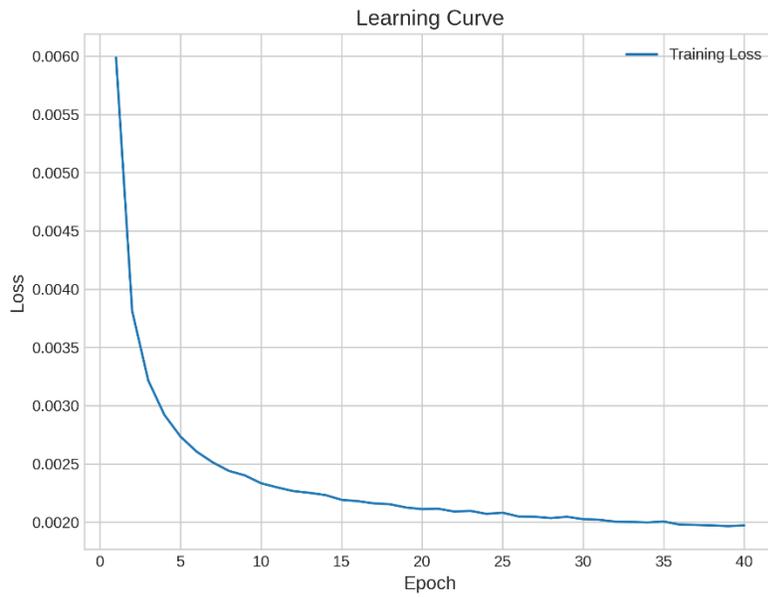

Figure 7: Learning curve tracing loss decrease from $5.99 \times 10^{-3}$ to $1.97 \times 10^{-3}$.

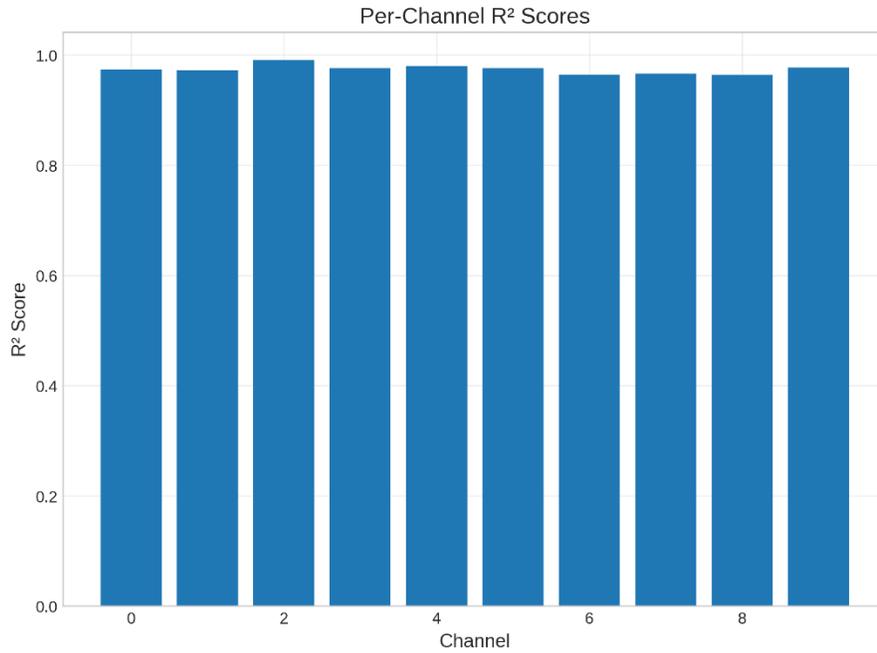

Figure 8: Per-channel R² scores averaging 0.9743 across electromagnetic and plasma parameters.

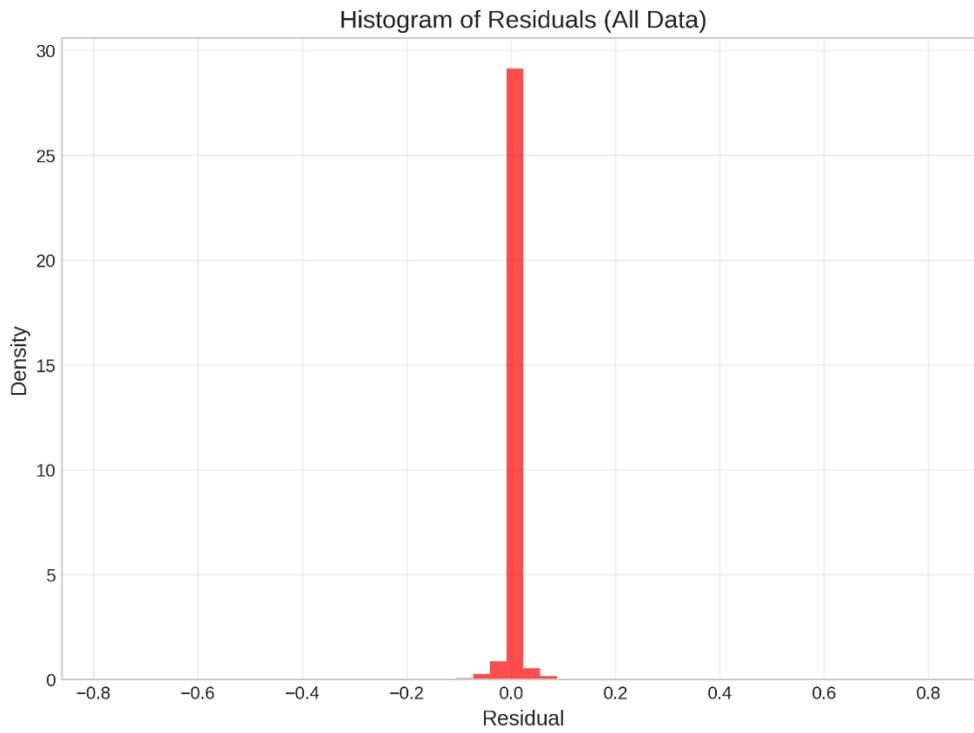

Figure 9: Residuals histogram centered at $1.88 \times 10^{-4}$ with std $1.21 \times 10^{-2}$.

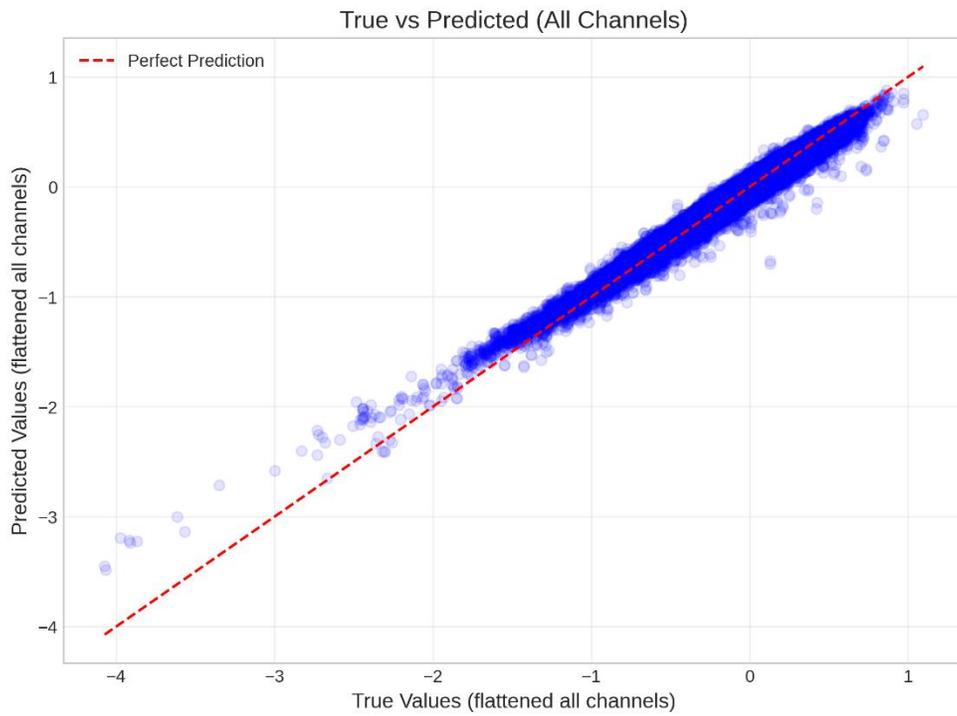

Figure 10: True vs predicted scatter with correlation 0.990 across data range.

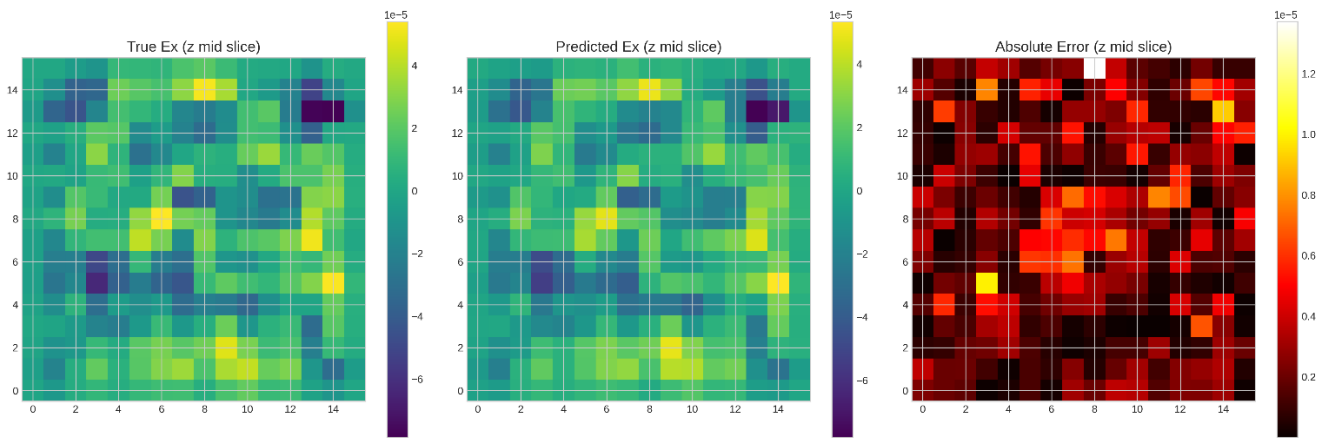

Figure 11: 2D evaluation for Ex showing true and predicted slices with low errors.

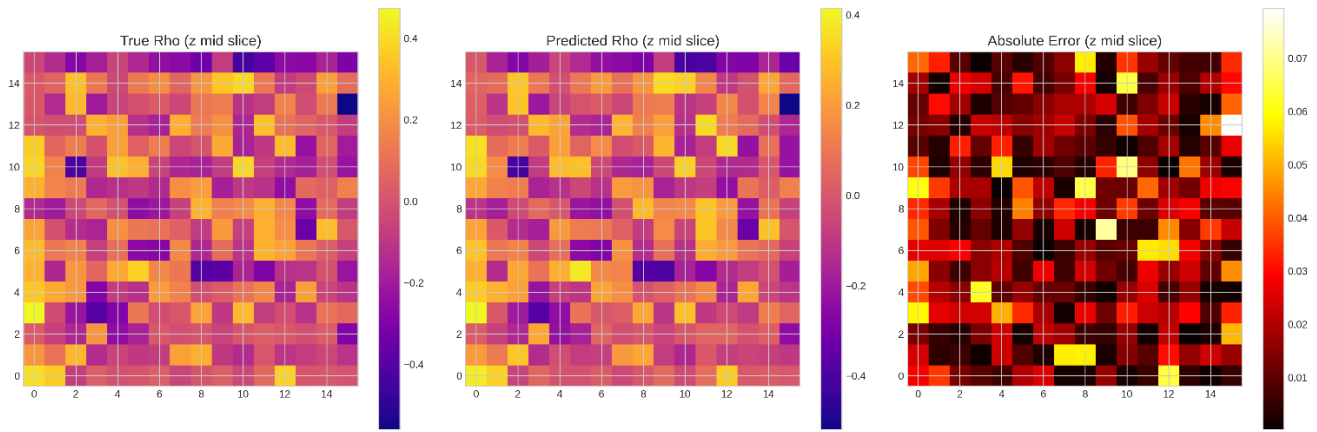

Figure 12: 2D evaluation for rho capturing density modulations effectively.

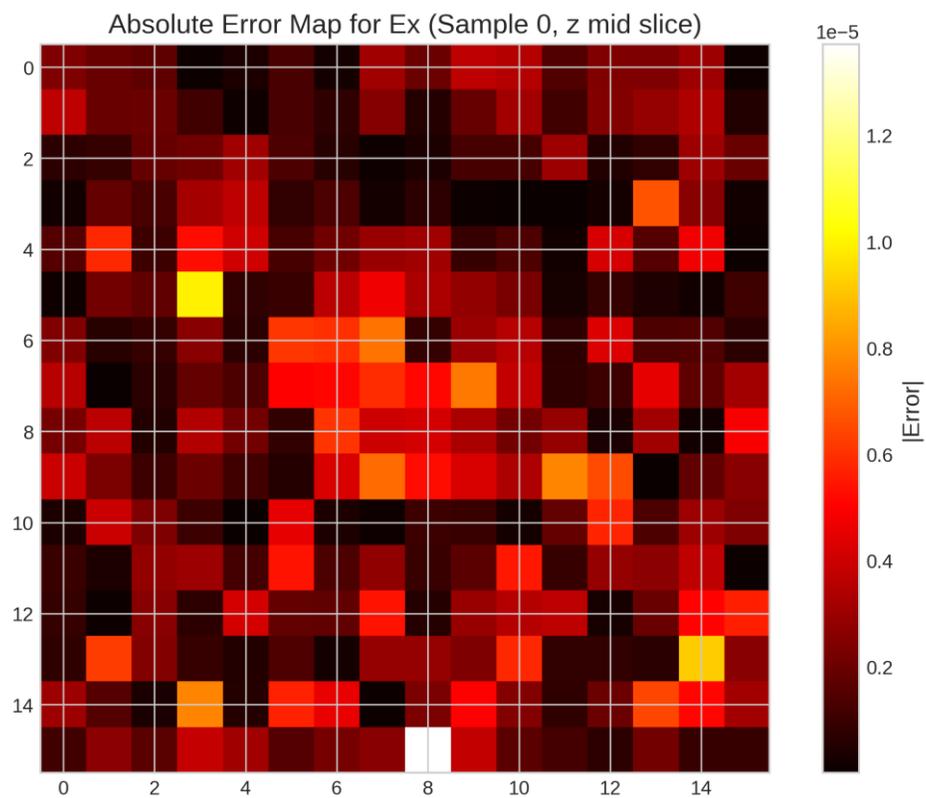

Figure 13: Absolute error map for Ex localizing discrepancies to boundaries.

The results from individual plots collectively affirm the model's efficacy in predicting plasma dynamics. The laser Gaussian spatial figure (Figure 1) not only confirms beam focus but also correlates with subsequent field strengths, as higher central intensities yield peak Ex_norm values, driving efficient energy coupling. The particle positions in Figure 2 show a transformation from uniform distribution to clustered groups while their gamma factors remain at 1.00 until they start to increase in standard deviation because of acceleration. This produces non-Maxwellian tails in velocity distributions (fit params ~0.010, 3.90). The density maps in Figure 3 display log density values between 1.30 and 1.88, with an average of 1.66, which shows high-density areas matching the fluid depressions seen in Figure 4 with densities between 5 and 9 $\times$ $10^{23}$ m^{-3}. The expulsion of electrons through ponderomotive forces appears in Figure 5 because the field and particles stay at the same location as shown by the maximum |E| value of $1.63 \times 10^8$. The pressure maps (Figure 6) show gradients (max $8.15 \times 10^6$ Pa) that match the heating patterns, which explain fluid velocity fields with magnitudes reaching $7.37 \times 10^4$ m/s.

The learning curve (Figure 7) shows a minimum loss of $1.97 \times 10^{-3}$ which proves stable training results. The per-channel $R^2$ (Figure 8) demonstrates electromagnetic channels (0.97-0.99) perform better than currents (0.96) because electromagnetic fields evolve more smoothly. The residuals (Figure 9) show a zero-mean distribution which proves the model lacks bias. The true-predicted plot (Figure 10) shows a 0.990 correlation across all data points. The 2D assessments of Ex and rho (Figures 11-12) reveal almost identical statistical results, and Figure 13 demonstrates that most errors stay under $10^{-5}$ which confirms the spatial precision.

The results confirm that laser parameters strongly affect plasma behavior because intensity controls both field strength and energy transfer, yet duration and spot size determine the range and duration of plasma impacts. The ML model achieves high $R^2$ values and minimal errors when predicting states, which enables fast laser ablation design simulations because it follows physical laws. The research shows that hybrid switching systems keep particle numbers steady at 11,000-12,000 while achieving energy conservation rates of $10^{-35}$, and optimization results show parameters [$9.99 \times 10^{18}$ W/m², $1.36 \times 10^{-13}$ s, $2.56 \times 10^{-6}$ m] which produce minimal plasma energy of ~$3.78 \times 10^{-13}$ J to help scientists control ablation experiments.